\pdfoutput=1
\documentclass[conference]{IEEEtran}
\IEEEoverridecommandlockouts
\usepackage{cite}
\usepackage{amsmath,amssymb,amsfonts}
\usepackage{algorithmic}
\usepackage{graphicx}
\usepackage{textcomp}
\usepackage{multirow}
\usepackage{booktabs}
\usepackage{float}  
\usepackage{caption}
\usepackage{colortbl} 
\usepackage{xcolor}
\definecolor{mymagenta}{RGB}{234,0,124} 
\usepackage{hyperref}
\hypersetup{
	colorlinks=true,
	linkcolor=red,
	filecolor=blue,
	urlcolor=mymagenta,
	citecolor=green
}
\def\BibTeX{{\rm B\kern-.05em{\sc i\kern-.025em b}\kern-.08em
    T\kern-.1667em\lower.7ex\hbox{E}\kern-.125emX}}
\begin{document}

\title{Visual Semantic Description Generation with MLLMs for Image-Text Matching}
\author{
	Junyu Chen, Yihua Gao and Mingyong Li$^{\dagger}$ \\
	Chongqing Normal University, Chongqing, China
	\\
	\{2023210516022, 2023210516033\}@stu.cqnu.edu.cn, 
	}
\maketitle
\renewcommand{\thefootnote}{}
\footnotetext{$^{\dagger}$ Corresponding author (email: limingyong@cqnu.edu.cn)}

\maketitle

\begin{abstract}
Image-text matching (ITM) aims to address the fundamental challenge of aligning visual and textual modalities, which inherently differ in their representations—continuous, high-dimensional image features vs. discrete, structured text. We propose a novel framework that bridges the modality gap by leveraging multimodal large language models (MLLMs) as visual semantic parsers. By generating rich Visual Semantic Descriptions (VSD), MLLMs provide semantic anchor that facilitate cross-modal alignment. 
Our approach combines: (1) Instance-level alignment by fusing visual features with VSD to enhance the linguistic expressiveness of image representations, and (2) Prototype-level alignment through VSD clustering to ensure category-level consistency. These modules can be seamlessly integrated into existing ITM models. Extensive experiments on Flickr30K and MSCOCO demonstrate substantial performance improvements. The approach also exhibits remarkable zero-shot generalization to cross-domain tasks, including news and remote sensing ITM. The code and model checkpoints are available at \href{https://github.com/Image-Text-Matching/VSD}{https://github.com/Image-Text-Matching/VSD}.
\end{abstract}


\section{Introduction}
\label{sec:intro}
Image-text matching is a core research task in the multimodal domain, aiming to bridge the heterogeneous gap between image and text to achieve efficient cross-modal alignment. As a foundational component of multimodal large language models (MLLMs) and text-to-image diffusion models, image-text matching (ITM) models play a crucial role in cross-modal understanding and generation.

The heterogeneous gap between image and text primarily manifests in the fundamental differences in information representation. Image consists of continuous high-dimensional pixel matrices, containing rich visual features such as color, brightness, and texture. However, these features are often sparsely distributed and highly redundant, with not all pixels being critical for semantic expression. In contrast, text comprises discrete symbol sequences, following specific grammatical rules and semantic structures, organizing information in a highly structured and linearized manner. Compared to image, textual information is densely concentrated with low redundancy, where each linguistic unit carries explicit semantics and can construct complete semantic expressions through ordered combinations. This difference in representation makes cross-modal alignment a challenging task.

To bridge this gap, researchers have introduced additional visual semantic parsers as inductive biases to effectively extract semantic information from image and facilitate image-text alignment. For instance, Faster R-CNN \cite{faster_r_cnn} has been widely applied to extract salient regions in image, thereby obtaining key semantics and filtering redundant information. Building upon this, SCAN \cite{SCAN} achieves fine-grained alignment between salient region features and textual tokens through stacked cross-attention mechanisms. VSRN++ \cite{VSRN++} further designs a semantic reasoning network to learn feature representations containing crucial scene concepts by parsing local information. Moreover, researchers have explored various visual semantic parsing strategies. For example, ACLIP \cite{ACLIP} combines attention mechanisms with EMA networks to filter image features highly relevant to the text while discarding low-relevance regions. CORA \cite{CORA} parses image captions into scene graphs, establishing relational connections between object and attribute nodes. These visual semantic parsing methods, which incorporate prior knowledge, have significantly improved image-text matching performance.

Methods represented by CLIP \cite{CLIP} and SigLip \cite{SigLip}, while not designed with dedicated visual semantic parsers, have achieved powerful generalization capabilities through large-scale image-text pair pretraining. However, these methods not only heavily rely on massive datasets but may have limitations in performance on specific tasks. Some studies have introduced additional visual semantic parsers to enhance the performance of pretrained models on specific tasks. RegionCLIP \cite{RegionCLIP} introduces a region-text alignment module on CLIP, effectively improving CLIP's performance in region recognition and object detection tasks. SCLIP \cite{SCLIP} proposes a correlative self-attention mechanism, enhancing the model's performance in semantic segmentation tasks.
\begin{figure*}[t]
	\centering
	\includegraphics[width=\textwidth]{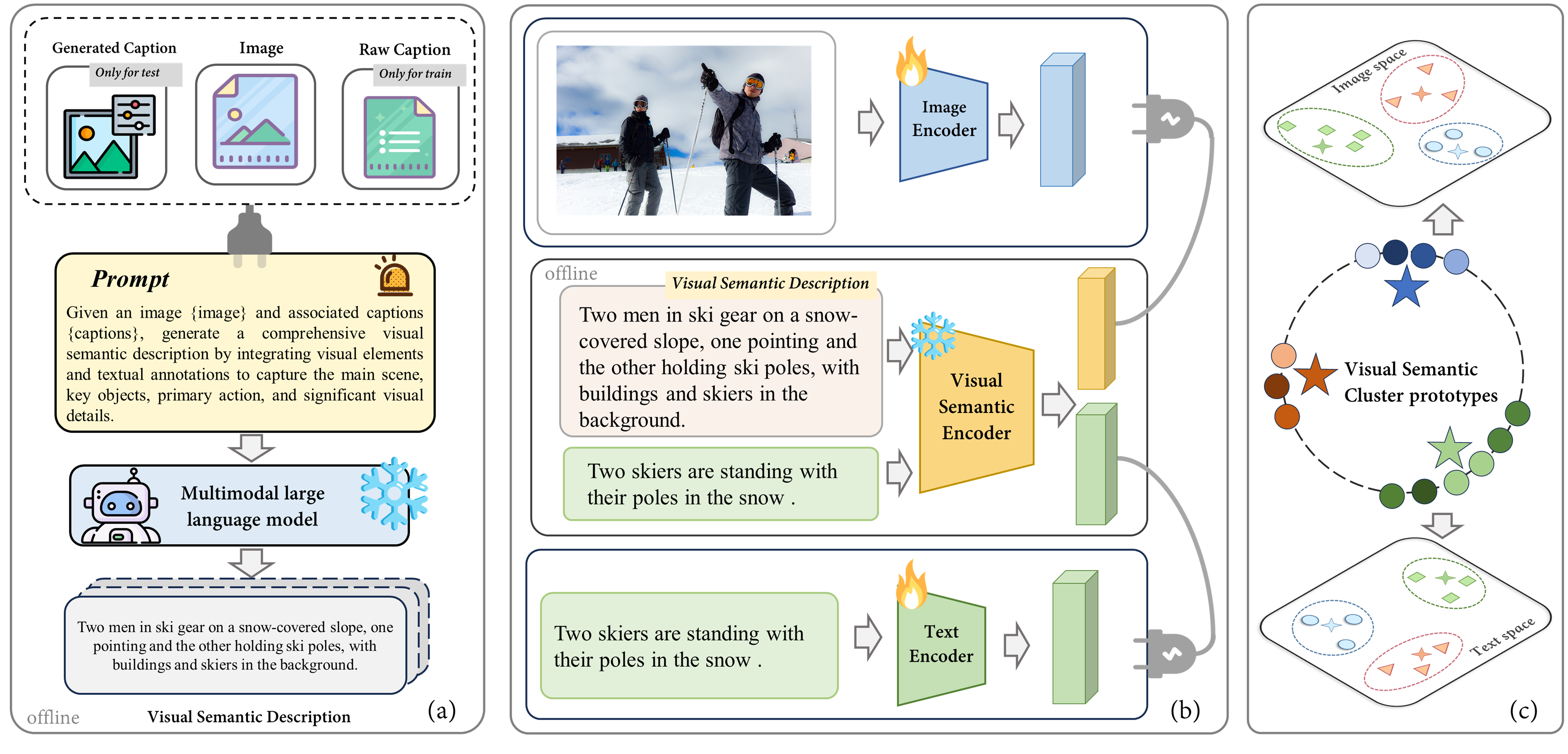}
	\caption{Illustration of the proposed framework. (a) \textbf{Visual Semantic Description Generation}: A MLLM is utilized to produce high-quality, information-rich visual semantic description  by integrating visual elements and textual annotations, capturing key objects, actions, and scene details. (b) \textbf{Instance-level Semantic Alignment Enhancement}: This module strengthens cross-modal semantic alignment by adaptively fusing raw image features with representations of the generated descriptions, enabling fine-grained instance-level alignment. (c) \textbf{Prototype-level Semantic Alignment Enhancement}: Semantic variations across samples are captured by clustering visual semantic descriptions, enabling cross-modal alignment at the prototype level within the clustered semantic space.
	} \label{fig:overview}
	\vspace{-1.0em}
\end{figure*}
MLLMs have demonstrated powerful visual semantic understanding capabilities. How to fully utilize these capabilities to enhance traditional image-text matching methods warrants in-depth exploration. Recent research has mainly focused on leveraging MLLMs to improve the quality of visual-language datasets: LaCLIP \cite{LaCLIP} uses ChatGPT to rewrite original image captions, thereby improving caption quality and diversity; Nguyen et al. \cite{nguyen2024improving} employe BLIP-2 \cite{BLIP2} to rewrite captions for image-text pairs with low matching scores in datasets; Liu et al. \cite{liu2023mllms} propose combining multiple MLLMs to collaboratively rewrite captions and ensure the quality and usability of extended captions through text pruning techniques. Unlike existing work, this paper explores using MLLMs as visual semantic parsers to enhance image-text alignment by generating high-quality and information-rich visual semantic descriptions. 

Our main contributions can be summarized as follows:
	\begin{itemize}
	\item
	We propose a new paradigm of using MLLMs as visual semantic parsers, effectively bridging the heterogeneous gap between image and text by generating high-quality visual semantic descriptions as semantic anchor.
	\item
	We design semantic alignment strategies at both instance and prototype levels: the former enhances cross-modal alignment of individual samples through adaptive feature fusion; the latter models semantic relationships between samples using clustering structures to achieve semantic consistency at the category level.
	\item 
	Experiments show that the proposed method can serve as a plug-and-play module to improve the performance of existing image-text matching models. It achieves significant performance improvements on Flickr30K and MSCOCO, while demonstrating excellent zero-shot generalization capabilities in cross-domain image-text matching scenarios, including news and remote sensing.
	\end{itemize}
\section{Methodology}
We provide a comprehensive overview of our methodology, as illustrated in Fig. \ref{fig:overview}.
\subsection{Preliminaries}
Image-Text Matching (ITM) aims to evaluate the semantic consistency between image and text. Given a dataset containing $N$ image-text pairs, denoted as ${(I_i, T_i)}_{i=1}^N$, where $(I_i, T_i)$ represents a semantically matched positive sample pair, and $(I_i, T_j)_{i \neq j}$ denotes semantically unrelated negative sample pairs.
Mainstream ITM approaches typically adopt a dual-stream architecture, projecting both image and text into a shared visual-semantic embedding space. Specifically, an image encoder $f(\cdot)$ encodes image $I_i$ into a feature vector $\mathbf{v}_i$, while a text encoder $g(\cdot)$ maps text $T_i$ to a feature vector $\mathbf{t}_i$. To achieve cross-modal semantic alignment, these methods often employ a hard negative triplet loss or a contrastive loss as the optimization objective. These losses aim to minimize the feature distance of positive sample pairs while maximizing that of negative sample pairs, thereby achieving semantic consistency across modalities.
\subsection{Visual Semantic Description Generation}
To enhance the model's semantic understanding capability for image-text matching, we utilize MLLM (specifically MiniCPMV-2.6\cite{MiniCPMV}, though our method is adaptable to any MLLM) to generate high-quality and information-rich visual semantic descriptions. Specifically, given an input image $I$ and its associated caption set $\mathcal{C} = \{c_1, c_2, \dots, c_k\}$, we design the following prompt:

\textit{Given an image \{image\} and associated captions \{captions\}, generate a comprehensive visual semantic description (15-30 words) by integrating visual elements and textual annotations to capture the main scene, key objects, primary action, and significant visual details.}

Based on this prompt, the generation process of the visual semantic description is formalized as:
\begin{equation}
	d_{\text{vsd}} = \text{MLLM}(\text{Prompt}, I, \mathcal{C}),
\end{equation}
where $d_{\text{vsd}}$ represents the generated visual semantic description.

Considering that caption information is unavailable during the testing phase, we design a two-stage generation strategy to ensure consistency in data distribution between training and testing phases:
\begin{itemize}
\item
Use MLLM to parse the input image $I$ and generate a set of semantically relevant captions $\hat{\mathcal{C}}$.
\item
Input the image $I$, the generated caption set $\hat{\mathcal{C}}$, and the prompt together into MLLM to obtain the final visual semantic description $d_{\text{vsd}}$.
\end{itemize}

The generated visual semantic description serves as a plug-and-play auxiliary information, significantly enhancing the semantic alignment capability of the image-text matching model. Specifically, we introduce BGE-large-v1.5 \cite{BGE}, an advanced model from the text retrieval domain, as the visual semantic encoder to map the visual semantic description into a high-dimensional semantic space:
\begin{equation}
	\mathbf{\O}_{\text{vsd}} = \text{BGE}(d_{\text{vsd}}) \in \mathbb{R}^{L \times d},
\end{equation}
where $\mathbf{\O}_{\text{vsd}}$ is the sequence representation of the visual semantic description, $L$ is the sequence length, and $d$ is the feature dimension. To obtain a compact semantic representation, we apply mean pooling to the sequence representation:
\begin{equation}
	\tilde{\mathbf{d}}_{\text{vsd}} = \text{MeanPool}(\mathbf{\O}_{\text{vsd}}) \in \mathbb{R}^d.
\end{equation}

This fully leverages the cross-modal understanding capability of MLLM, providing an additional semantic bridge for the dual-stream architecture and enhancing the semantic alignment between image and text. Meanwhile, to ensure semantic consistency, we use the same visual semantic encoder to encode the original text $T_i$, obtaining the corresponding feature representation $\tilde{\mathbf{t}}_i$.
\subsection{Instance-level Semantic Alignment Enhancement}
After obtaining the visual semantic description representation $\tilde{\mathbf{d}}_{\text{vsd}}$, we design an instance-level semantic alignment mechanism to enhance the model's cross-modal semantic alignment capability through adaptive fusion of the original image feature and the visual semantic description representation. To accommodate different image-text matching methods and dataset distribution characteristics, we design an adaptive feature fusion strategy based on a gating mechanism.

Specifically, for the image feature $\mathbf{v}$, we introduce a gating variable $g$ to control its fusion ratio with the visual semantic description representation $\tilde{\mathbf{d}}_{\text{vsd}}$:
\begin{equation}
	\begin{gathered}
		\hat{\mathbf{v}} = g \cdot \mathbf{v} + (1 - g) \cdot \tilde{\mathbf{d}}_{\text{vsd}}, \\
		g = \operatorname{sigmoid}(\mathbf{W}_v [\mathbf{v}, \tilde{\mathbf{d}}_{\text{vsd}}] + b_v),
	\end{gathered}
\end{equation}
where $\mathbf{W}_v$ and $b_v$ are learnable parameters.

For text features, we first obtain the feature vector $\mathbf{t}$ through the text encoder for the original text $T_i$, while simultaneously deriving $\tilde{\mathbf{t}}$ via the visual semantic encoder. Subsequently, we fuse the two using a similar gating mechanism:
\begin{equation}
	\begin{gathered}
		\hat{\mathbf{t}} = h \cdot \mathbf{t} + (1 - h) \cdot \tilde{\mathbf{t}}, \\
		h = \operatorname{sigmoid}(\mathbf{W}_t [\mathbf{t}, \tilde{\mathbf{t}}] + b_t),
	\end{gathered}
\end{equation}

Based on the obtained fused representations $\hat{\mathbf{v}}$ and $\hat{\mathbf{t}}$, we construct an Instance-level Semantic Alignment (ISA) loss to enhance the semantic alignment of cross-modal features. This loss takes the form of a hard negative triplet:
\begin{equation}
	\mathcal{L}^\text{ISA}=[\gamma-s(\boldsymbol{\hat v},\boldsymbol {\hat t})+s(\boldsymbol{\hat v},\,\boldsymbol{\hat t^-})]_++[\gamma-s(\boldsymbol{\hat v},\boldsymbol {\hat t})+s(\boldsymbol{\hat v^-},\boldsymbol{\hat t})]_+,
\end{equation}
where $\gamma$ is the margin, $s(\cdot,\cdot)$ represents the similarity measurement function, $[\cdot]_+$ denotes the positive operation, and $\boldsymbol{\hat t^-}$ and $\boldsymbol{\hat v^-}$ are the hardest negative text and image in the batch, respectively.

The visual semantic description serves as a bridge to facilitate this instance-level alignment mechanism, which enriches image semantics through feature fusion while strengthening cross-modal semantic associations.
\subsection{Prototype-level Semantic Alignment Enhancement}
To better exploit the semantic structure within visual semantic descriptions, we introduce a prototype-level alignment mechanism. We first capture semantic differences between samples through K-means clustering, and then achieve cross-modal alignment in the clustered semantic space, fully utilizing the inherent semantic structure information in the data.

Specifically, we perform K-means clustering on the visual semantic description representations $\tilde{\mathbf{d}}_{\text{vsd}} \in \mathbb{R}^{m \times d}$ to generate a set of prototypes $\mathbf{P} = {\{\mathbf{p}_1, \mathbf{p}_2, \dots, \mathbf{p}_k\}} \in \mathbb{R}^{d \times k}$, where $k$ is the number of prototypes. Each prototype represents a low-dimensional projection of visual semantic descriptions, storing frequent and common feature patterns that can serve as universal representations to enhance various image-text matching methods.

For image embeddings $\mathbf{V} = {\{\mathbf{v}_1, \dots, \mathbf{v}_m\}} \in \mathbb{R}^{d \times m}$ and text embeddings $\mathbf{T} = {\{\mathbf{t}_1, \dots, \mathbf{t}_m\}} \in \mathbb{R}^{d \times m}$, we project them into a unified prototype space to obtain their respective classification scores $\mathbf{u}^{v}$ and $\mathbf{u}^{t}$:
\begin{equation}
	\begin{gathered}\mathbf{u}^{v} = \operatorname{softmax}(\mathbf{V}^\top \mathbf{P}) \in \mathbb{R}^{m \times k}, \\\mathbf{u}^{t} = \operatorname{softmax}(\mathbf{T}^\top \mathbf{P}) \in \mathbb{R}^{m \times k},
	\end{gathered}
\end{equation}
\begin{table*}[tbp]
	\small
	\setlength{\tabcolsep}{1.8mm}
	\centering
	\caption{Image-text retrieval  performance comparison on Flickr30K and MSCOCO. $^\star$ and $^\dagger$ denote visual semantic description generated by Florence-2-large-ft-0.77B and MiniCPMV2.6-8B respectively, while * indicates ensemble model results.}
	\label{tab:main_result}
	\begin{tabular}{lcccccccccccccc} 
		\toprule
		\multirow{3}{*}{Methods} & \multicolumn{7}{c}{Flickr30K (1K Test Set)}                                                                              & \multicolumn{7}{c}{MSCOCO (5K Test Set)}                                                                            \\
		& \multicolumn{3}{c}{Image-to-Text}             & \multicolumn{3}{c}{Text-to-Image}             & \multirow{2}{*}{rSum} & \multicolumn{3}{c}{Image-to-Text}             & \multicolumn{3}{c}{Text-to-Image}             & \multirow{2}{*}{rSum}  \\
		& R@1  & R@5  & R@10  & R@1  & R@5 & R@10  &                       & R@1  & R@5  & R@10  & R@1  & R@5 & R@10          &                        \\ 
		\midrule
		SCAN* \cite{SCAN} & 67.4 & 90.3 & 95.8 & 48.6 & 77.7 & 85.2 & 465.0 & 50.4 & 82.2 & 90.0 & 38.6 & 69.3 & 80.4 & 410.9 \\
		VSRN++ \cite{VSRN++} & 79.2 & 94.6 & 97.5 & 60.6 & 85.6 & 91.4 & 508.9 & 54.7 & 82.9 & 90.9 & 42.0 & 72.2 & 82.7 & 425.4 \\
		CHAN \cite{CHAN} & 80.6 & 96.1 & 97.8 & 63.9 & 87.5 & 92.6 & 518.5 & 59.8 & 87.2 & 93.3 & 44.9 & 74.5 & 84.2 & 443.9 \\
		CORA* \cite{CORA} & 83.4 & 95.9 & 98.6 & 64.1 & 88.1 & 93.1 & 523.3 & 64.3 & 87.5 & 93.6 & 45.4 & 74.7 & 84.6 & 450.1 \\
		IMEB \cite{IMEB} & 84.2 & 96.7 & 98.4 & 64.0 & 88.0 & 92.8 & 524.1 & 62.8 & 87.8 & 93.5 & 44.9 & 74.6 & 84.0 & 447.6 \\
		GPO \cite{GPO} & 81.7 & 95.4 & 97.6 & 61.4 & 85.9 & 91.5 & 513.5 & 58.3 & 85.3 & 92.3 & 42.4 & 72.7 & 83.2 & 434.3 \\ \addlinespace[1.5pt]
		\rowcolor[rgb]{0.942,0.968,1}  ~ + \textit{VSD}$^\star$ & 83.1 & 97.1 & \textbf{99.0} & 66.2 & 88.9 & 93.6 & 527.9 & 61.9 & 86.1 & 92.7 & 45.1 & 74.5 & 83.8 & 444.2 \\
		\rowcolor[rgb]{0.942,0.968,1} ~ + \textit{VSD}$^\dagger$ & \textbf{86.1} & \textbf{97.9} & 98.5 & \textbf{71.9} & \textbf{91.6} & \textbf{94.8} & \textbf{540.8} & \textbf{65.1} & \textbf{88.1} & \textbf{93.5} &\textbf{49.9} & \textbf{76.6} & \textbf{85.1} & \textbf{458.2} \\
		HREM \cite{HREM} & 83.3 & 96.0 & 98.1 & 63.5 & 87.1 & 92.4 & 520.4 & 61.8 & 87.0 & 93.2 & 44.0 & 73.7 & 83.4 & 443.1 \\ \addlinespace[1.5pt]
		\rowcolor[rgb]{0.942,0.968,1} ~ + \textit{VSD}$^\star$ & 83.9 & 97.2 & \textbf{98.9} & 66.8 & 89.0 & 93.6 & 529.5 & 63.8 & 87.4 & 93.6 & 45.2 & 74.7 & 84.0 & 448.7 \\
		\rowcolor[rgb]{0.942,0.968,1} ~ + \textit{VSD}$^\dagger$ & \textbf{86.7} & \textbf{97.7} & \textbf{98.9} & \textbf{72.5} & \textbf{91.4} & \textbf{94.9} & \textbf{542.1} & \textbf{66.7} & \textbf{88.5} & \textbf{94.2} & \textbf{50.0} & \textbf{76.4} & \textbf{85.1} & \textbf{460.7} \\ 
		CLIP$_\text{ViT-B/32}$ \cite{CLIP} & 86.3 & 96.7 & 98.8 & 69.9 & 90.1 & 94.3 & 536.1 & 60.1 & 83.8 & 90.9 & 45.3 & 72.9 & 82.1 & 435.1 \\ \addlinespace[1.5pt]
		\rowcolor[rgb]{0.942,0.968,1} ~~ + \textit{VSD}$^\star$ & 88.5 & \textbf{98.7} & \textbf{99.8} & 75.0 & 93.4 & 96.1 & 551.5 & 65.2 & 87.5 & 93.7 & 49.3 & 76.5 & 85.3 & 457.6 \\ 
		\rowcolor[rgb]{0.942,0.968,1} ~~ + \textit{VSD}$^\dagger$ & \textbf{90.2} & 98.6 & 99.7 & \textbf{77.9} & \textbf{93.9} & \textbf{96.4} & \textbf{556.7} & \textbf{67.8} & \textbf{88.7} & \textbf{94.1} & \textbf{51.5} & \textbf{77.8} & \textbf{85.8} & \textbf{465.8} \\
		\bottomrule
	\end{tabular}
	\vspace{-1.0em}
\end{table*}

Then, we use the Sinkhorn algorithm \cite{Sinkhorn} to obtain soft assignment matrices $\mathbf{D}^v, \mathbf{D}^t \in \mathbb{R}^{m \times k}$ for the classification scores. Based on this, we define the Prototype-level Semantic Alignment (PSA) loss:
\begin{equation}
\begin{aligned}
	\mathcal{L}_{\text{img}}^{\text{PSA}} &= -\frac{1}{m} \sum_{i=1}^m \sum_{j=1}^k \mathbf{D}^t_{ij} \log \frac{\exp(\mathbf{u}^v_{ij} / \tau)}{\sum_{l=1}^k \exp(\mathbf{u}^v_{il} / \tau)}, \\
	\mathcal{L}_{\text{txt}}^{\text{PSA}} &= -\frac{1}{m} \sum_{i=1}^m \sum_{j=1}^k \mathbf{D}^v_{ij} \log \frac{\exp(\mathbf{u}^t_{ij} / \tau)}{\sum_{l=1}^k \exp(\mathbf{u}^t_{il} / \tau)},
\end{aligned}
\end{equation}
where $\tau$ is a temperature coefficient used to control the smoothness of the distribution. The final prototype-level semantic alignment loss is defined as:
\begin{equation}
	\mathcal{L}^{\text{PSA}} = \mathcal{L}_{\text{img}}^{\text{PSA}} + \mathcal{L}_{\text{txt}}^{\text{PSA}}.
\end{equation}
\subsection{Objective Function}
The final objective function is:
\begin{equation}
	\mathcal{L} = \mathcal{L}^{\text{PSA}} + 	\mathcal{L}^\text{ISA}.
\end{equation}
\section{experiment}
\subsection{Datasets and Evaluation Metrics}
This study evaluates the proposed method on two benchmark datasets: Flickr30K and MSCOCO. Each image in both datasets is associated with five descriptive captions. Flickr30K consists of 31,783 images, following the standard split protocol \cite{splittingprotocol} with 29,000 images for training, 1,000 for validation, and 1,000 for testing. The MSCOCO dataset comprises 123,287 images, split into 113,287 training images, 5,000 validation images, and 5,000 test images, consistent with previous work \cite{GPO}. Additionally, to assess the cross-domain image-text matching generalization of the model, we introduce the N24News \cite{N24news} dataset from the news domain and the RSITMD \cite{RSITMD} dataset from the remote sensing domain for zero-shot testing. Performance evaluation employs standard retrieval metrics R@K (K=1,5,10) and rSum, where rSum is the sum of R@K values for bidirectional image-text retrieval.
\subsection{Implementation Details}
We validate our method using three models: GPO \cite{GPO}, HREM \cite{HREM}, and CLIP \cite{CLIP} (OpenCLIP ViT-B/32), with GPO and HREM retaining their original configurations. CLIP is fine-tuned using hard negative triplet loss (Adam optimizer, lr=5e-7).
For semantic description generation, we use MiniCPMV2.6-8B \cite{MiniCPMV} with two-stage prompting and Florence-2-large-ft-0.77B \cite{Florence-2} for direct captioning. Both models remain untuned and can be easily deployed on a single NVIDIA RTX 3090 GPU. Descriptions are encoded via BGE-large-v1.5 \cite{BGE}.  Our method is implemented with PyTorch on NVIDIA RTX A6000. 
Training uses batch sizes of 128 (Flickr30k) and 256 (MSCOCO) for 25 epochs. The number of prototypes $k$ is set to 896 and 2560 respectively, with the prototype semantic alignment temperature $\tau=0.1$.
\vspace{-0.2em}
\subsection{Main Results}
As shown in Tab. \ref{tab:main_result}, we conducted comprehensive experimental evaluations on two benchmark datasets, Flickr30K and MSCOCO. The experimental results demonstrate that the proposed Visual Semantic Description (VSD)  enhances image-text alignment method significantly improves the performance of various baselines. Specifically, we employed two different models to generate visual semantic descriptions: the Image Captioning model Florence-2-large-ft-0.77B (denoted as VSD$^\star$) and the MLLM model MiniCPMV2.6-8B (denoted as VSD$^\dagger$).

On the Flickr30K test set, applying VSD$^\dagger$ to the GPO baseline improves the Image-to-Text R@1 metric from 81.7\% to 86.1\%, Text-to-Image R@1 metric from 61.4\% to 71.9\%, and rSum from 513.5 to 540.8. Similarly, on the MSCOCO test set, the R@1 improves by 6.8\% for Image-to-Text and 7.5\% for Text-to-Image, with the overall rSum increasing by 23.9\%. When VSD$^\dagger$ is applied to the CLIP model, the performance gains are even more pronounced, significantly outperforming existing baseline methods.

It is worth noting that, although VSD$^\dagger$ generally provides larger performance improvements, showcasing the strong cross-modal understanding capabilities of MLLM, VSD$^\star$ also delivers substantial performance improvement. Considering the lightweight nature of the Florence-2-large-ft-0.77B model, VSD$^\star$ offers greater practicability in resource-constrained scenarios, providing a flexible option for real-world deployment.
\subsection{Generalization Capability}
\begin{table}[tbp]
	\centering
	\caption{Comparison of model performance trained on MSCOCO and evaluated on Flickr30K test set.}
	\label{tab:corss_dataset}
	\resizebox{\columnwidth}{!}{
		\begin{tabular}{@{}l@{\hspace{4pt}}ccccccc@{}} 
			\toprule
			\multirow{3}{*}{\small Methods} & \multicolumn{3}{c}{\small Image-to-Text}  & \multicolumn{3}{c}{\small Text-to-Image} & \multirow{2}{*}{\small rSum} \\
			\cmidrule(lr){2-4} \cmidrule(lr){5-7}
			& R@1 & R@5 & R@10 & R@1 & R@5 & R@10 & \\
			\midrule
			CHAN & 68.7 & 91.5 & 95.3 & 55.6 & 81.2 & 87.5 & 479.8 \\
			IMEB & 72.3 & 90.2 & 95.1 & 54.8 & 80.0 & 86.9 & 479.3 \\
			GPO & 68.0 & 89.2 & 93.7 & 50.0 & 77.0 & 84.9 & 462.8 \\ \addlinespace[1pt]
			\rowcolor[rgb]{0.942,0.968,1} ~ + \textit{VSD}$^\dagger$ & 84.5 & 97.1 & 98.5 & 69.8 & 89.4 & 93.8 & 533.1 \\
			HREM & 70.3 & 90.7 & 95.2 & 52.9 & 78.7 & 86.2 & 474.0 \\ \addlinespace[1pt]
			\rowcolor[rgb]{0.942,0.968,1} ~ + \textit{VSD}$^\dagger$ & 85.1 & 97.2 & 98.6 & 69.9 & 89.7 & 93.7 & 534.2 \\
			CLIP$_\text{ViT-B/32}$ & 82.5 & 95.3 & 98.3 & 67.1 & 88.4 & 92.8 & 524.4 \\ \addlinespace[1pt]
			\rowcolor[rgb]{0.942,0.968,1} ~ + \textit{VSD}$^\dagger$ & \textbf{89.5} & \textbf{98.7} & \textbf{99.2} & \textbf{75.7} & \textbf{92.7} & \textbf{95.8} & \textbf{551.6} \\
			\bottomrule
		\end{tabular}%
	}
\end{table}
\begin{table}[tbp]
	\vspace{-1.0em}
	\centering
	\caption{Performance comparison of zero-shot cross-domain image-text retrieval on news domain (N24News) and remote sensing domain (RSITMD) datasets.}
	\label{tab:cross_domain}
	\resizebox{\columnwidth}{!}{%
		\begin{tabular}{@{}l@{\hspace{8pt}}l@{\hspace{4pt}}ccccccc@{}} 
			\toprule
			\multirow{3}{*}{\small Dataset} & \multirow{3}{*}{\small Methods} & \multicolumn{3}{c}{\small Image-to-Text}  & \multicolumn{3}{c}{\small Text-to-Image} & \multirow{2}{*}{\small rSum} \\
			\cmidrule(lr){3-5} \cmidrule(lr){6-8}
			& & R@1 & R@5 & R@10 & R@1 & R@5 & R@10 & \\
			\midrule
			\multirow{4}{*}{{N24News}} 
			& ALBEF \cite{ALBEF} & 21.2 & 38.1 & 46.2 & 21.5 & 38.3 & 46.0 & 211.1 \\
			& BLIP2 \cite{BLIP2} & 33.0 & 53.6 & 61.4 & 32.3 & 53.5 & 62.2 & 296.0 \\
			& CLIP$_\text{ViT-B/32}$ \cite{CLIP} & 48.3 & 70.0 & \textbf{77.1} & 43.3 & 64.4 & 71.9 & 374.9 \\
			\rowcolor[rgb]{0.942,0.968,1}
			& ~ + \textit{VSD}$^\dagger$ & \textbf{58.5} & \textbf{71.0} & 75.6 & \textbf{59.3} & \textbf{72.3} & \textbf{76.8} & \textbf{413.5} \\
			\midrule
			\multirow{4}{*}{{RSITMD}}
			& GaLR \cite{GaLR} & 14.8 & 31.6 & 42.5 & 11.2 & 36.7 & 51.7 & 188.5 \\
			& HVSA \cite{HVSA} & 13.2 & 32.1 & \textbf{45.6} & 11.4 & 39.2 & \textbf{57.5} & 198.9 \\
			& CLIP$_\text{ViT-B/32}$ \cite{CLIP} & 9.5 & 23.0 & 32.7 & 8.8 & 27.9 & 43.2 & 145.1 \\
			\rowcolor[rgb]{0.942,0.968,1}
			& ~ + \textit{VSD}$^\dagger$ & \textbf{18.4} & \textbf{34.7} & 44.2 & \textbf{18.5} & \textbf{41.1} & 55.5 & \textbf{212.4} \\
			\bottomrule
		\end{tabular}%
	}
	\vspace{-2.0em}
\end{table}
To comprehensively evaluate the generalization capability of our proposed method, we conducted experimental validations from two perspectives: cross-dataset transfer and cross-domain transfer.

A common concern is that incorporating MLLMs may increase the risk of model overfitting to specific datasets. To evaluate cross-dataset generalization, we conducted zero-shot transfer experiments by training on MSCOCO and testing on Flickr30K. As shown in Tab. \ref{tab:corss_dataset}, our Visual Semantic Description (VSD) enhancement strategy significantly improves cross-dataset performance. For GPO, incorporating VSD$^\dagger$ improved the rSum by 70.3\% (462.8 to 533.1). Similar substantial improvements were observed for HREM (60.2\%) and CLIP (27.2\%). These consistent improvements across different model architectures indicate our method's effectiveness in enhancing cross-dataset generalization and capturing universal vision-language features.

We further evaluated cross-domain transfer on N24News (news image) and RSITMD (remote sensing image) datasets, which exhibit substantial domain differences from the training data. As shown in Tab. \ref{tab:cross_domain}, on N24News, our method outperforms CLIP baseline with R@1 improvements of 10.2\% and 16\% for Image-to-Text and Text-to-Image retrieval, respectively. Similar improvements were observed on RSITMD despite its larger domain shift.

These experimental results indicate that by incorporating visual semantic description generated by MLLMs, our method not only achieves effective transfer within similar domains but demonstrates the ability to handle scenarios with significant domain shifts, providing a viable solution for improving the generalization performance of image-text matching methods.
\subsection{Ablation Study}
\begin{table}[tbp]
	\centering
	\caption{Ablation study of different components in our method based on the GPO baseline, reporting the average results of image-to-text and text-to-image retrieval.}
	\label{tab:Ablation}
	\resizebox{\columnwidth}{!}{
		\begin{tabular}{@{}l@{\hspace{4pt}}cccccc@{}} 
			\toprule
			\multirow{3}{*}{\small Methods} & \multicolumn{3}{c}{ Flickr30K}  & \multicolumn{3}{c}{MSCOCO} \\
			\cmidrule(lr){2-4} \cmidrule(lr){5-7}
			& R@1 & R@5 & R@10 & R@1 & R@5 & R@10 \\
			\midrule
			GPO & 68.0 & 89.2 & 93.7 & 50.0 & 77.0 & 84.9 \\ 		
			~ + \textit{PSA} & 73.1 & 91.8 & 95.7 & 52.2 & 80.2 & 88.3 \\
			~ + \textit{VSD}$^\star$ & 74.7 & 93.0 & 96.3 & 53.5 & 80.3 & 88.3 \\
			~ + \textit{VSD}$^\dagger$(w/o \textit{PSA}) & 78.4 & 93.8 & 96.3 & 57.2 & 81.5 & 88.7 \\
			~ + \textit{VSD}$^\dagger$ & \textbf{79.0} & \textbf{94.8} & \textbf{96.7} & \textbf{57.5} & \textbf{82.4} & \textbf{89.3} \\
			\bottomrule
		\end{tabular}
	}
	\vspace{-1.5em}
\end{table}
To evaluate the impact of each module on model performance, we progressively incorporate different components into the GPO baseline and conduct comprehensive experiments on both Flickr30K and MSCOCO test sets. Tab. \ref{tab:Ablation} presents the average performance metrics for Image-to-Text (I2T) and Text-to-Image (T2I) retrieval tasks.
\subsubsection{Effectiveness of Visual Semantic Description}
The experimental results demonstrate that incorporating the Visual Semantic Description (VSD) module into the GPO baseline leads to substantial performance improvements. Specifically, ~ + \textit{VSD}$^\dagger$(w/o \textit{PSA}) achieves a 10.4\% increase in R@1 on Flickr30K (from 68.0\% to 78.4\%) and a 7.2\% improvement on MSCOCO (from 50.0\% to 57.2\%) compared to the baseline. These results indicate that the semantically rich visual description generated by MLLM effectively bridge the modality gap by transforming visual information into content-rich structured language representations, significantly enhancing cross-modal alignment.
\subsubsection{Effectiveness of Prototype-level Semantic Alignment Enhancement}
The Prototype-level Semantic Alignment Enhancement (PSA) module also demonstrates significant performance contributions. + \textit{PSA} achieves consistent improvements across all metrics compared to the baseline. By performing semantic clustering on VSD, the PSA module obtains high-quality prototype references, guiding ITM to achieve accurate cross-modal alignment in the clustered semantic space while leveraging the inherent structural information in the data.
\subsubsection{Comparison of Visual Semantic Description Generation Strategies}
By comparing + \textit{VSD}$^\star$ (using Florence-2-large-ft-0.77B) with + \textit{VSD}$^\dagger$ (using MiniCPMV2.6-8B), we find that visual semantic description generated by the MLLM approach yield superior performance. This indicates that MLLM has greater advantages in bridging the semantic gap between modalities. However, considering the lightweight nature of Florence-2-large-ft-0.77B, it may offer higher practical value in real-world deployment scenarios.
\begin{figure}[t]
	\centering
	\includegraphics[width=\linewidth]{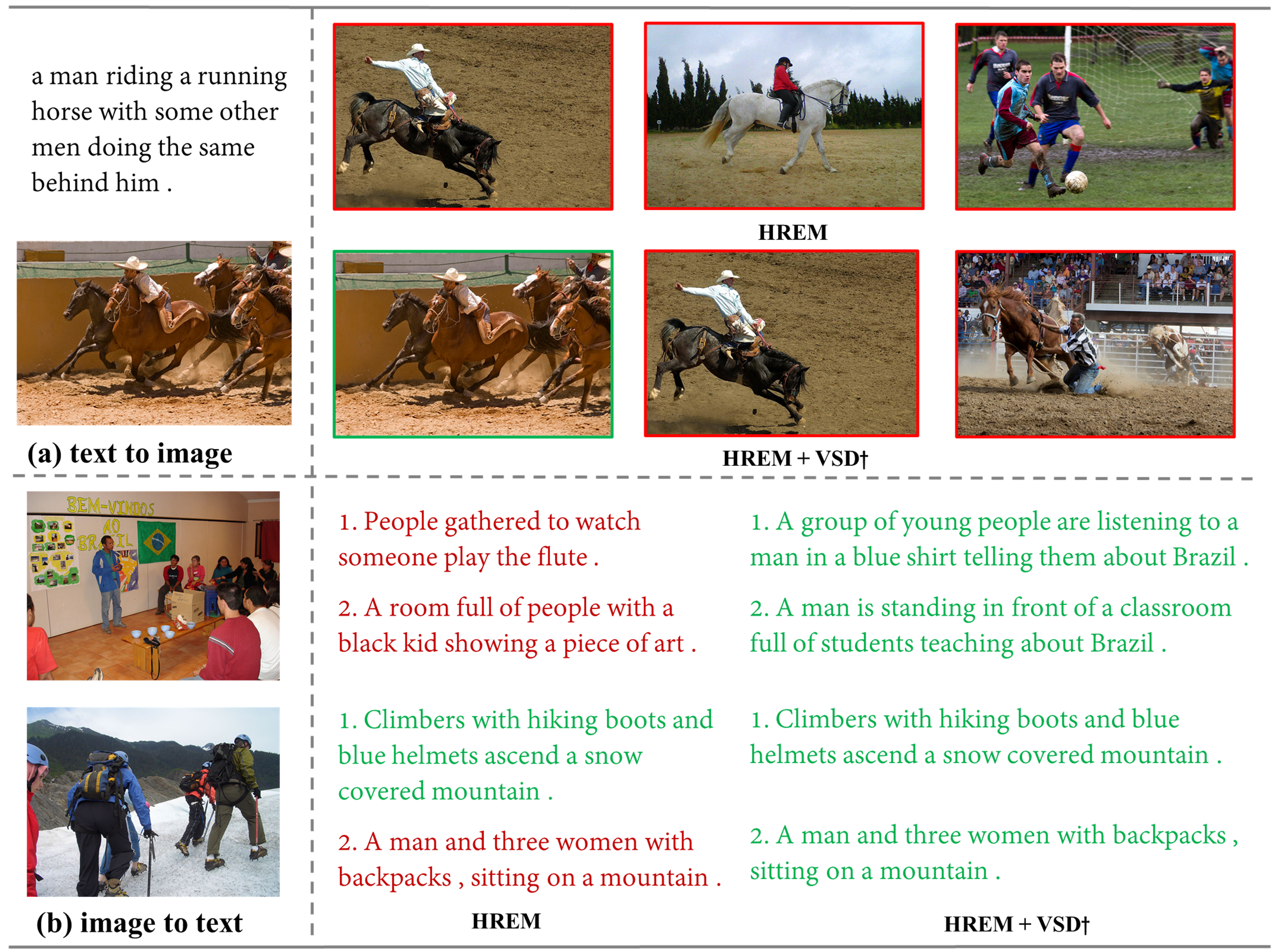}
	
	\caption{ Qualitative comparison results of bidirectional retrieval on the Flickr30K test set. All models were trained on the MSCOCO dataset and evaluated on Flickr30K test set for cross-dataset assessment. The correct text retrieval results are in green color and the incorrect ones are in red color.}
	\label{fig:retrieve_case_study}
	\vspace{-1.0em}
\end{figure}
\subsection{Qualitative Analysis}
To better demonstrate the effectiveness of our method, we adopted a cross-dataset evaluation strategy by applying models trained on MSCOCO to the Flickr30K test set. We selected representative samples from the Flickr30K test set for visualization analysis. As shown in Fig. \ref{fig:retrieve_case_study}, our proposed HREM+VSD$^\dagger$ method demonstrates significant advantages over the baseline HREM in cross-dataset bidirectional retrieval tasks. Notably, our method exhibits superior capability in capturing semantic alignment between image and text. This performance demonstrates strong generalization ability.
\section{Conclusion}
This paper presents a framework utilizing multimodal large language models (MLLMs) as visual semantic parsers to address the modality gap in image-text matching (ITM). The approach generates visual semantic descriptions that serve as semantic anchor, facilitating cross-modal alignment through complementary instance-level and prototype-level alignment strategies. 
Extensive experiments on Flickr30K and MSCOCO benchmarks demonstrates superior performance, while zero-shot evaluations on cross-domain tasks validate the robust generalization capability of our method. These results highlight the potential of integrating MLLMs to advance ITM performance and cross-modal understanding. Future work will focus on two directions: integrating more powerful MLLMs to enhance visual semantic descriptions, and extending our approach to broader scenarios such as video-text alignment and multimodal document retrieval. 
\bibliographystyle{IEEEbib}
\bibliography{icme2025references}
\end{document}